# Evolution of superconductivity in isovalent Te-substituted $K_xFe_{2-y}Se_2$ crystals

Toshinori Ozaki,[1] Hiroyuki Takeya,[1] Keita Deguchi,[1,2] Satoshi Demura,[1,2] Hiroshi Hara,[1,2] Tohru Watanabe,[1,2] Saleem James Denholme,[1] Hiroyuki Okazaki,[1] Masaya Fujioka,[1] Yoshiko Yokota,[1] Takahide Yamaguchi,[1] and Yoshihiko Takano[1,2]

[1] *National Institute for Materials Science, 1-2-1 Sengen, Tsukuba, Ibaraki 305-0047, Japan*

[2] *University of Tsukuba, 1-1-1Tennodai, Tsukuba, Ibaraki 305-0001, Japan*

(Abstract)

We report the evolution of superconductivity and the phase diagram of the $K_xFe_{2-y}Se_{2-z}Te_z$ ($z$ = 0 – 0.6) crystals grown by a one-step synthesis. The one-step synthesis is much simpler than a conventional process to grow $K_xFe_{2-y}Se_2$ system. No structural transition is observed in any crystals, while lattice parameters exhibit a systematic expansion with increasing Te content. The $T_c$ exhibits a gradual decrease with increasing Te content from $T_c^{onset}$ = 32.9 K at $z$ = 0 to $T_c^{zero}$ = 27.9 K at $z$ = 0.5, followed by a sudden suppression of superconductivity at $z$ = 0.6. Upon approaching a Te concentration of 0.6, the shielding volume fraction decreases and eventually drops to zero. Simultaneously, hump positions in $\rho$-$T$ curve shift to lower temperatures. These results suggest that isovalent substitution of Te for Se in $K_xFe_{2-y}Se_2$ crystals suppresses the superconductivity in this system.

I. INTRODUCTION

The recent discovery of the alkaline-intercalated iron-selenide superconductor $A_yFe_{2-x}Se_2$ (A = K, Cs, Rb,)[1-4] with a transition temperature $T_c$ of about 30 K have brought new excitement to the field of iron-based superconductors. Unlike other iron-based superconductors which are metals with spin-density-wave (SDW) order[5,6], the superconductivity in this class is in the proximity of an antiferromagnetic (AFM) Mott insulator[7,8], similar to the cuprate high-temperature superconductors. Both the angle-resolved photoemission (ARPES) experiments[9,10] and the band calculations[11] showed that the Fermi surface has only electron pockets, while the hole bands sink below the Fermi level. The early results of muon-spin rotation ($\mu$SR)[12,13], neutron diffraction[14], resistivity and magnetic investigations[15] have revealed a coexistence of superconductivity and a strong antiferromagnetic order, with an extremely large magnetic moment of 3.31 $\mu_B$ on the iron atom and high Néel temperature of 559 K. Many experimental results revealed that the phase separation between the AFM and the superconducting phase occurred on nanoscopic length scales, and the superconducting phase does not have any Fe vacancies.[16-23]

Isovalent substitution could be effective to understand the correlation between structural parameters and physical properties in terms of chemical pressure, since it does not introduce additional electrons or holes into the system. As is well known, the PbO-type FeSe superconductor shows a great pressure effect on $T_c$. The $T_c$ of FeSe can reach 37 K under physical pressure from $T_c$ ~10 K[24,25], and can also increase up to 15 K by substituting Se with isovalent Te[26-28], corresponding to the negative pressure due to the fact that ionic radius of Te is larger than that of Se. These results imply that $T_c$ for FeSe can be strongly correlated with structural parameters. In the $A_xFe_{2-y}Se_2$ system, application of physical pressure showed that the $T_c$ was slightly increased, and then superconductivity was completely suppressed by further applied pressure.[29-33] However, there are hardly any reports on the negative chemical pressure achieved by isovalent substitution in the $A_xFe_{2-y}Se_2$ system.[34] This may be due to the difficulty and complexity in growing these materials. We have reported the successful growth of $K_xFe_{2-y}Se_2$ single crystals with high $J_c$ using a simple one-step process at relatively lower temperatures.[35] Here, we present the systematic investigation of structural parameters and physical properties in Te-substituted $K_xFe_{2-y}Se_2$ crystals grown by a one-step synthesis with an anticipation of the enhancement of $T_c$ by applying the negative chemical pressure.

## II. EXPERIMENTAL

Single crystals of $K_xFe_{2-y}Se_{2-z}Te_z$ ($z$ = 0 – 0.6) were grown by a simple one-step synthesis in the following way. Fe (99.9%), $K_2Se$ (99%) powders and Se (99.999%) and Te (99.999%) grains were put into an alumina crucible and sealed into an evacuated quartz tube. The mixture was slowly heated to 900°C and held for 3 hours. The melting mixture was, then, cooled down to 700°C at a rate of 4°C/h, followed by cooling down to room temperature by shutting off the furnace. The as-grown single crystals were sealed into quartz tube under vacuum and annealed at 400°C for 1 hour, followed by quenching in air.[35-37]

The obtained crystals were characterized by x-ray diffraction (XRD) with Cu-K$\alpha$ radiation at room temperature. The actual atomic composition of the crystals was determined by using energy dispersive x-ray spectrometry (EDX). At least five spots for each crystal have been measured to obtain the average composition. The measurement of resistivity was performed on a physical property measurement system (PPMS, Quantum Design). Magnetic susceptibility was measured using a superconducting quantum interference device (SQUID) magnetometer.

## III. RESULT AND DISCUSSION

All the obtained $K_xFe_{2-y}Se_{2-z}Te_z$ (z = 0–0.6) crystals have dark shiny surfaces. In order to confirm whether or not the Te is really incorporated into the $K_xFe_{2-y}Se_2$ compound, we investigated an actual composition of Te-substituted $K_xFe_{2-y}Se_2$ crystal with EDX analysis. The actual chemical composition for the crystals is given in Table 1. For all crystals, the detected Fe content in a unit cell is less than the nominal composition, indicative of the existence of Fe vacancies. It should be noted that the actual compositions of the Se and Te concentration in these crystals are close to nominal compositions. This result indicates that our growth method is well suited to grow $K_xFe_{2-y}Se_{2-z}Te_z$ crystals.

Figure 1(a) shows the x-ray diffraction patterns for $K_xFe_{2-y}Se_{2-z}Te_z$ ($z$ = 0–0.6) crystals. All the diffraction peaks can be indexed with the space group of *I*4/*m*, demonstrating that no structural transition can be detected. As presented in the inset of Fig. 1(a), the peak position of (246/426) shifts smoothly to a lower angle with increasing Te content. The lattice parameters *a*, *c* and a unit cell volume are displayed in Fig. 1(b) – 1(d), respectively. The lattice constants *a* and *c* calculated using the peak positions increase with increasing Te concentration due to the larger ionic size of $Te^{2-}$ than $Se^{2-}$, suggesting that the Te properly substitutes for Se in the $K_xFe_{2-y}Se_{2-z}Te_z$ system. We also found that the data points of the cell volume nearly fell into a line, approximately in accordance with Vegard's law.

Figure 2 shows the temperature dependence of magnetic susceptibility for $K_xFe_{2-y}Se_{2-z}Te_z$ crystals. The transition temperature $T_c^{mag}$ is listed in Table 1. Similar to observations in the $Rb_xFe_{2-y}Se_{2-z}Te_z$,[34] $T_c^{mag}$ was suppressed by the isovalent substitution of Te for Se. However, $K_xFe_{2-y}Se_{2-z}Te_z$ shows a more gradual decrease of $T_c$ up to Te-substitution of 0.5, then superconductivity was completely suppressed at $z$ = 0.6. Simultaneously, the shielding volume fraction decreases with increasing Te concentration. This may come from the increase of the AFM insulating phases with increasing Te-substitution in $K_xFe_{2-y}Se_{2-z}Te_z$. As mentioned above, there exists a phase separation between the superconducting and AFM insulating phases, which could take place at the mesoscopic scale in this system. Very recently, STM measurements revealed that the $K_xFe_{2-y}Se_2$ thin films contains four phases: the parent compound $KFe_2Se_2$, superconducting $KFe_2Se_2$ with √2×√5 charge ordering, superconducting $KFe_2Se_{2-z}$ with Se vacancies, and insulating $K_2Fe_4Se_5$ with √5×√5 Fe vacancy order.[23] We consider that the non-superconducting region such as the parent compound $KFe_2Se_2$ and the insulating $K_2Fe_4Se_5$ expand due to the increase of the Te-substitution for Se, and then the shielding volume fraction decrease.

Figure 3(a) displays the temperature dependence of the in-plane resistance for the $K_xFe_{2-y}Se_{2-z}Te_z$ crystals. The superconducting transition occurs at 32.9 K and reaches zero resistance at 32.1 K in the $K_{0.76}Fe_{1.68}Se_2$ crystal. As shown in the inset of Fig.3 (a), both $T_c^{onset}$ and $T_c^{zero}$ are systematically suppressed with increasing Te-substitution up to $z$ = 0.5, and the superconductivity disappears at $z$ = 0.6 (Fig. 3(b)). The values of $T_c^{onset}$ and $T_c^{zero}$ are given in Table 1. The $T_c$ behavior with gradual Te-substitution is in good agreement with the result of $Rb_xFe_{2-y}Se_{2-z}Te_z$,[34] especially with the steep decrease of $T_c^{onset}$ from 27.9 K at $z$ = 0.5 to zero at $z$ = 0.6. In $K_xFe_{2-y}Se_{2-z}Te_z$, however, a more continuous decrease of $T_c^{zero}$ up to 15.6 K at $z$ = 0.5 is observed. Such a difference may be due to the difference of the quality of the crystals, attributed to the different growth processes. Our growth process makes it possible to grow the $K_xFe_{2-y}Se_{2-z}Te_z$ crystals under a relatively low temperature of 900°C, which could suppress the evaporation of K atoms. For $z$ = 0.6, the resistance increases gradually with the decrease of temperature roughly following the thermally activated nature of semiconductors: $\rho = \rho_0\exp(E_a/k_BT)$, where $E_a$ is the activation energy.

Another remarkable feature of these compounds is the fact that its resistance exhibits a hump, showing a crossover from semiconducting behavior to metallic behavior at $T^{hump}$, and the hump shift to lower temperature with Te content. $T^{hump}$ seems to be correlated to superconductivity in this system. However, the recent high-pressure measurements on $Rb_xFe_{2-y}Se_2$[31] and $K_xFe_{2-y}Se_2$[32,33] suggest that the temperature of the hump is not related to superconductivity. Ying et al. report that resistivity hump could arise from the deficiency in Fe and Se in the conducting layers.[32] This indicates that the resistivity hump is related to the non-superconducting phase, since this phase, which has a deficiency in Fe, does not show superconductivity. Guo et al. proposed that the hump feature may result from a competition between the semiconducting state and the metallic state in the superconducting samples.[33] From these above reports and the result of the magnetization measurement in Fig. 2, it would be understood that the hump temperature is not correlated with superconductivity, but the ratio of the superconducting region to the semiconducting region caused by the phase separation in this system, i.e. the increase of the semiconducting region would bring about the shift of the hump position to a lower temperature.

We present the magnetic and superconducting phase diagram of $K_xFe_{2-y}Se_{2-z}Te_z$ in Fig. 4. It is obvious that both $T_c^{onset}$ and $T_c^{zero}$ exhibit a gradual decrease with increasing Te concentration and disappears with a Te-substitution of 0.6. Simultaneously, $T^{hump}$ decreases with Te content up to $z$ = 0.5. When the Te concentration comes to 0.6, only semiconducting behavior is observed without any metallic crossover.

## IV. CONCLUSION

In summary, we present the effect of the partial substitution of Te for Se on the physical properties and structural parameters in $K_xFe_{2-y}Se_{2-z}Te_z$ ($z$ = 0 – 0.6) crystals grown by a simple one-step synthesis. As expected, lattice parameters *a*, *c* and a unit cell volume exhibited a systematic expansion with increasing substitution of Te for Se. Upon approaching $z$ = 0.5, the superconducting $T_c$ is gradually suppressed, and suddenly vanishes at $z$ = 0.6, corresponding to the 30% of Te substitution. We found that the $T^{hump}$ shifts to lower temperatures with an increase of Te substitution, and simultaneously the shielding volume fraction also shows a decrease to approximately 10% at $z$ = 0.6. This result indicates that the hump position could be correlated with the proportion of superconducting phase to semiconducting phase formed by the phase separation. Our results testify that chemical substitution of Te for Se in $K_xFe_{2-y}Se_2$ crystal ultimately gives rises to the deterioration of superconductivity and the decrease of the superconducting phase by transference to semiconducting phases.

“Acknowledgments”

This work was supported in part by the Japan Society for the Promotion of Science (JSPS) through Grants-in-Aid for JSPS Fellows and ‘Funding program for World-Leading Innovative R&D on Science Technology (FIRST) Program’. This research was supported by Strategic International Collaborative Research Program (SICORP), Japan Science and Technology Agency.

(Captions)

Fig. 1 (a) X-ray diffraction patterns for $K_xFe_{2-y}Se_{2-z}Te_z$ ($z$ = 0 – 0.6) crystals at room temperature and fit using the $I4/m$ space group. Inset: (246/426) peak position with different Te concentrations. (b),(c) The lattice parameters $a$ and $c$ as a function of Te contetnts. (d) Te-substitution dependence of a unit cell volume in $K_xFe_{2-y}Se_{2-z}Te_z$ crystals.

Fig. 2 Temperature dependence of magnetic susceptibility for both zero-field cooling (ZFC) and field cooling (FC) procedures in a magnetic field of 20 Oe for $K_xFe_{2-y}Se_{2-z}Te_z$ ($z$ = 0 – 0.6) crystals.

Fig. 3 (a)Temperature dependence of the electric resistance for $K_xFe_{2-y}Se_{2-z}Te_z$ ($z$ = 0–0.5) crystals at 0 T. The inset enlarges resistivity curve at low temperature. (b) Resistance as a function of temperature for $K_xFe_{2-y}Se_{1.4}Te_{0.6}$ crystals at zero magnetic field.

Fig.4 Electric phase diagram showing $T_c^{onset}$, $T_c^{zero}$, and $T^{hump}$ as a function of $z$ in $K_xFe_{2-y}Se_{2-z}Te_z$.

Table 1. The nominal and actual composition, magnetic and transport properties of $K_xFe_{2-y}Se_{2-z}Te_z$ crystal. The actual compositions were characterized by EDX analysis. $T^{hump}$, $T_c$onset and $T_c^{zero}$ were obtained from resistivity curve. $T_c^{mag}$ was estimated by magnetization measurement.

| Nominal composition | Actual composition | $T^{hump}$ | $T_c^{onset}$ | $T_c^{zero}$ | $T_c^{mag}$ |
|---|---|---|---|---|---|
| $K_{0.8}Fe_2Se_2$ | $K_{0.78}Fe_{1.68}Se_2$[35] | 250 | 32.9 | 32.1 | 31.2 |
| $K_{0.8}Fe_2Se_{1.9}Te_{0.1}$ | $K_{0.686}Fe_{1.778}Se_{1.905}Te_{0.095}$ | 210 | 31.2 | 30.0 | 29.4 |
| $K_{0.8}Fe_2Se_{1.8}Te_{0.2}$ | $K_{0.735}Fe_{1.722}Se_{1.805}Te_{0.195}$ | 177 | 30.4 | 28.3 | 28.7 |
| $K_{0.8}Fe_2Se_{1.7}Te_{0.3}$ | $K_{0.710}Fe_{1.682}Se_{1.694}Te_{0.306}$ | 141 | 29.5 | 25.9 | 26.9 |
| $K_{0.8}Fe_2Se_{1.65}Te_{0.35}$ | $K_{0.724}Fe_{1.663}Se_{1.656}Te_{0.344}$ | 96 | 28.2 | 24.9 | 25.0 |
| $K_{0.8}Fe_2Se_{1.5}Te_{0.5}$ | $K_{0.717}Fe_{1.740}Se_{1.473}Te_{0.527}$ | 94 | 27.9 | 16.7 | 14.9 |
| $K_{0.8}Fe_2Se_{1.4}Te_{0.6}$ | $K_{0.749}Fe_{1.766}Se_{1.371}Te_{0.629}$ | — | — | — | — |

Figure 1(a)

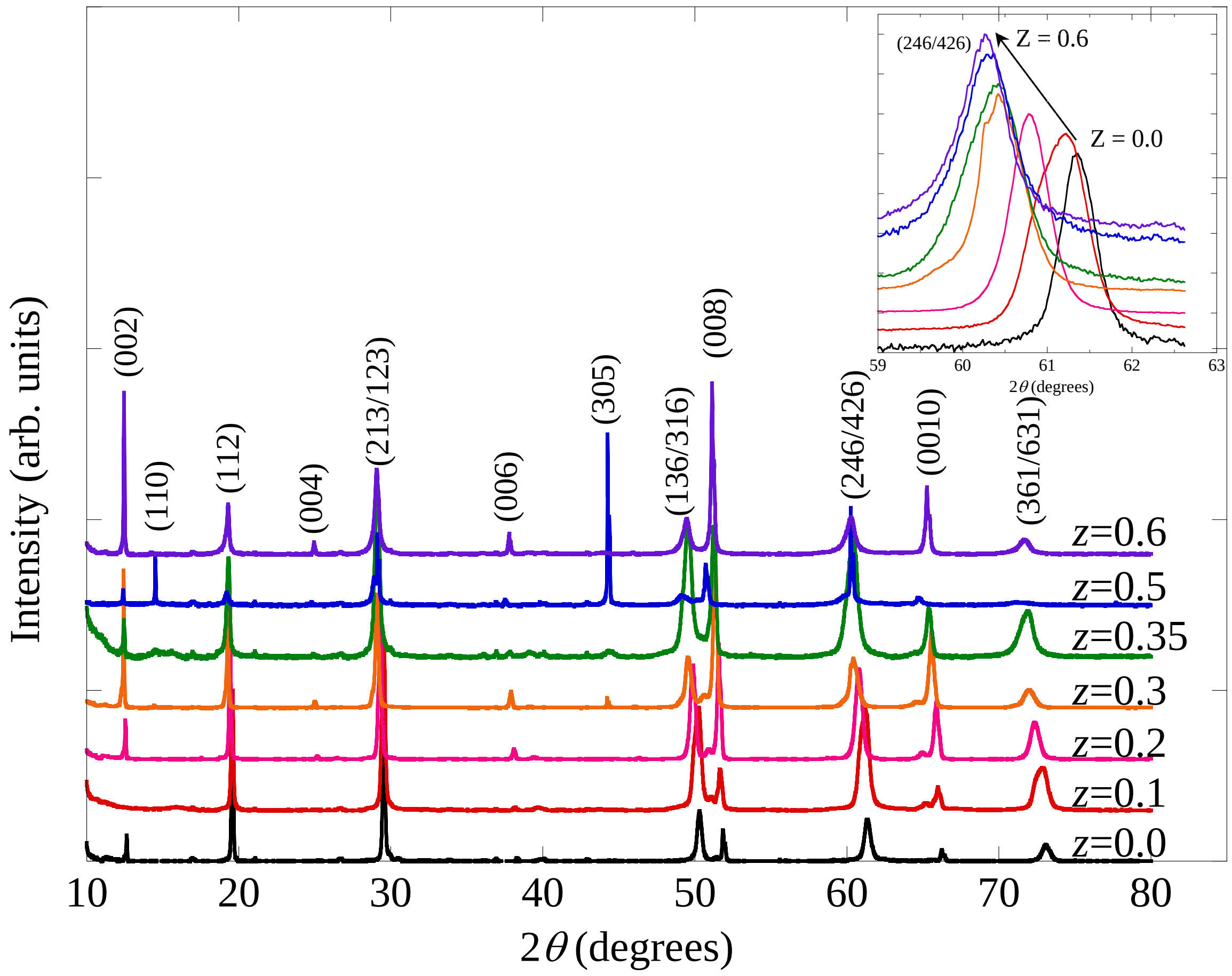

Figure 1(b)-(d)

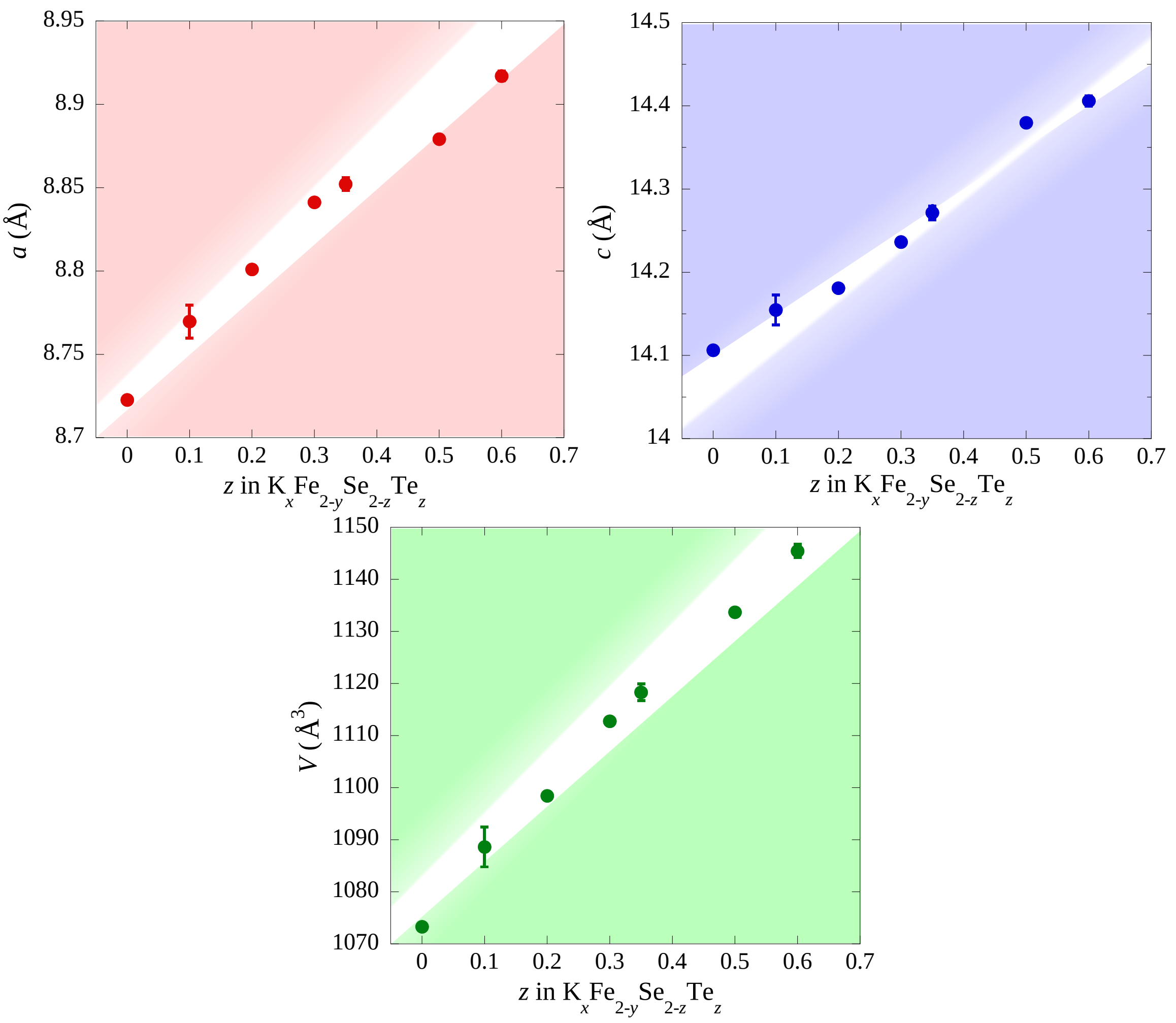

8.95
8.9
8.85
8.8
8.75
8.7
a (Å)
0 0.1 0.2 0.3 0.4 0.5 0.6 0.7
z in KxFe2-ySe2-zTez
14.5
14.4
14.3
14.2
14.1
14
c (Å)
0 0.1 0.2 0.3 0.4 0.5 0.6 0.7
z in KxFe2-ySe2-zTez
1150
1140
1130
1120
1110
1100
1090
1080
1070
V (Å3)
0 0.1 0.2 0.3 0.4 0.5 0.6 0.7
z in KxFe2-ySe2-zTez

Figure 2

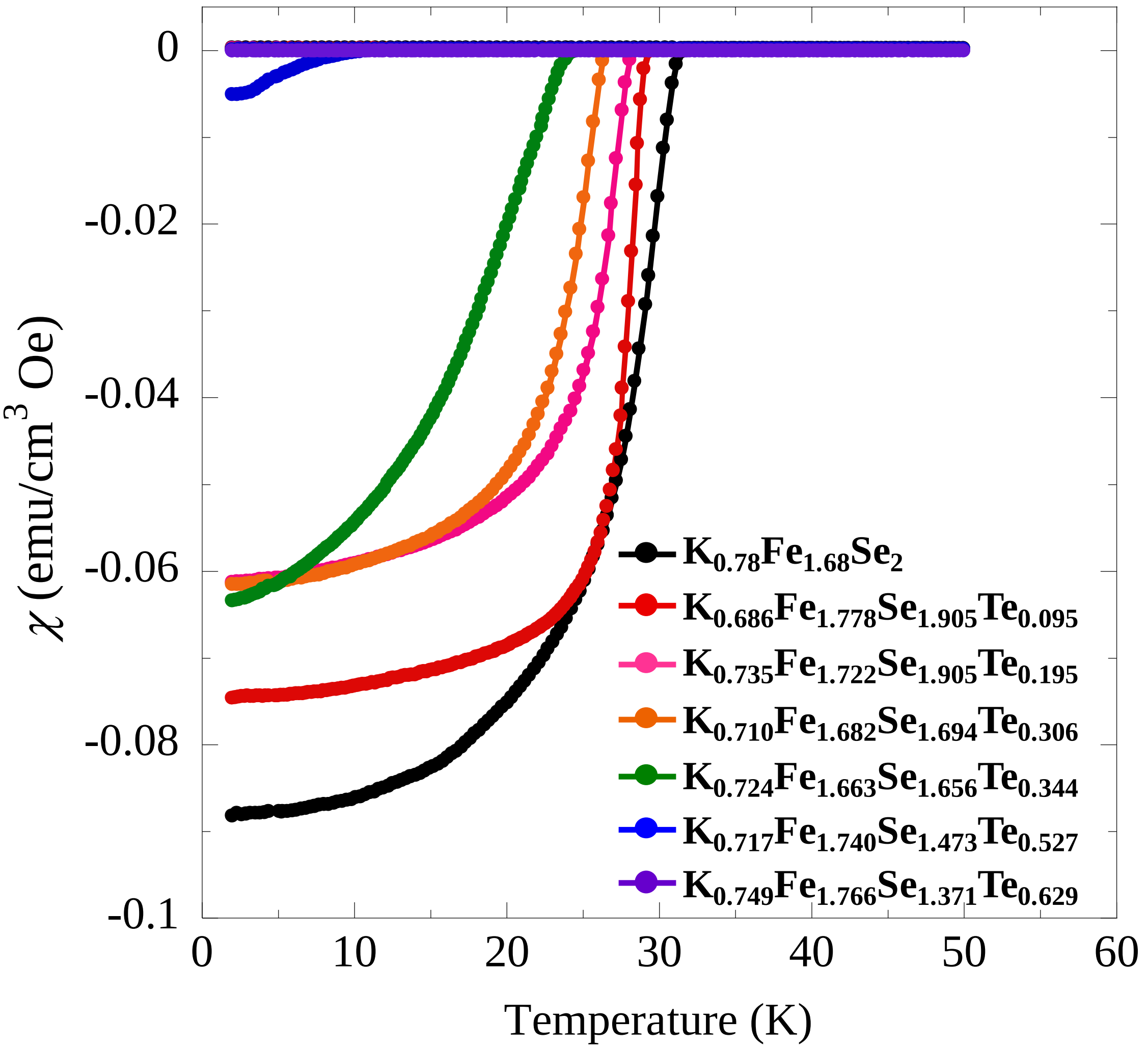

Figure 3(a)

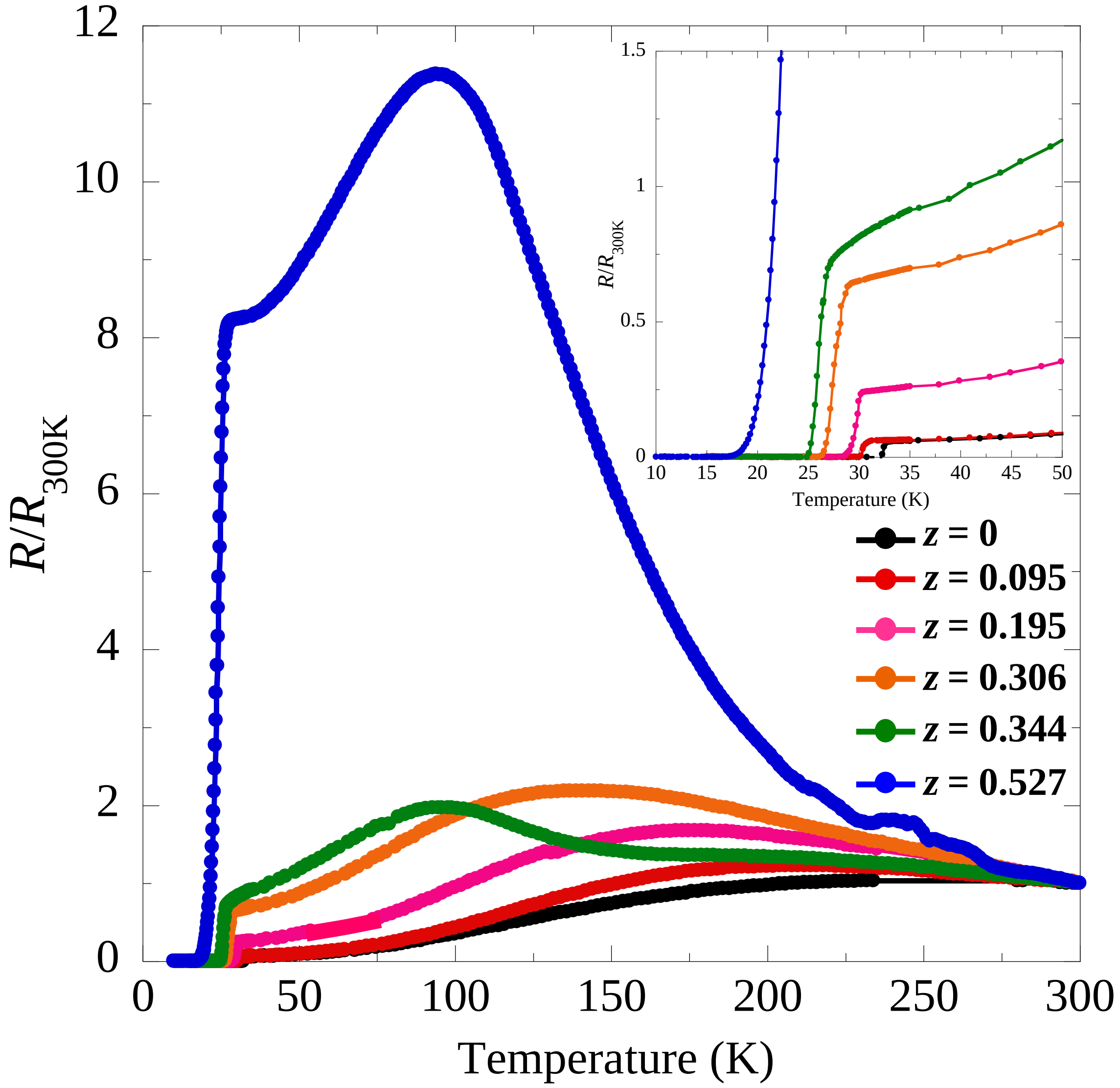

Figure 3(b)

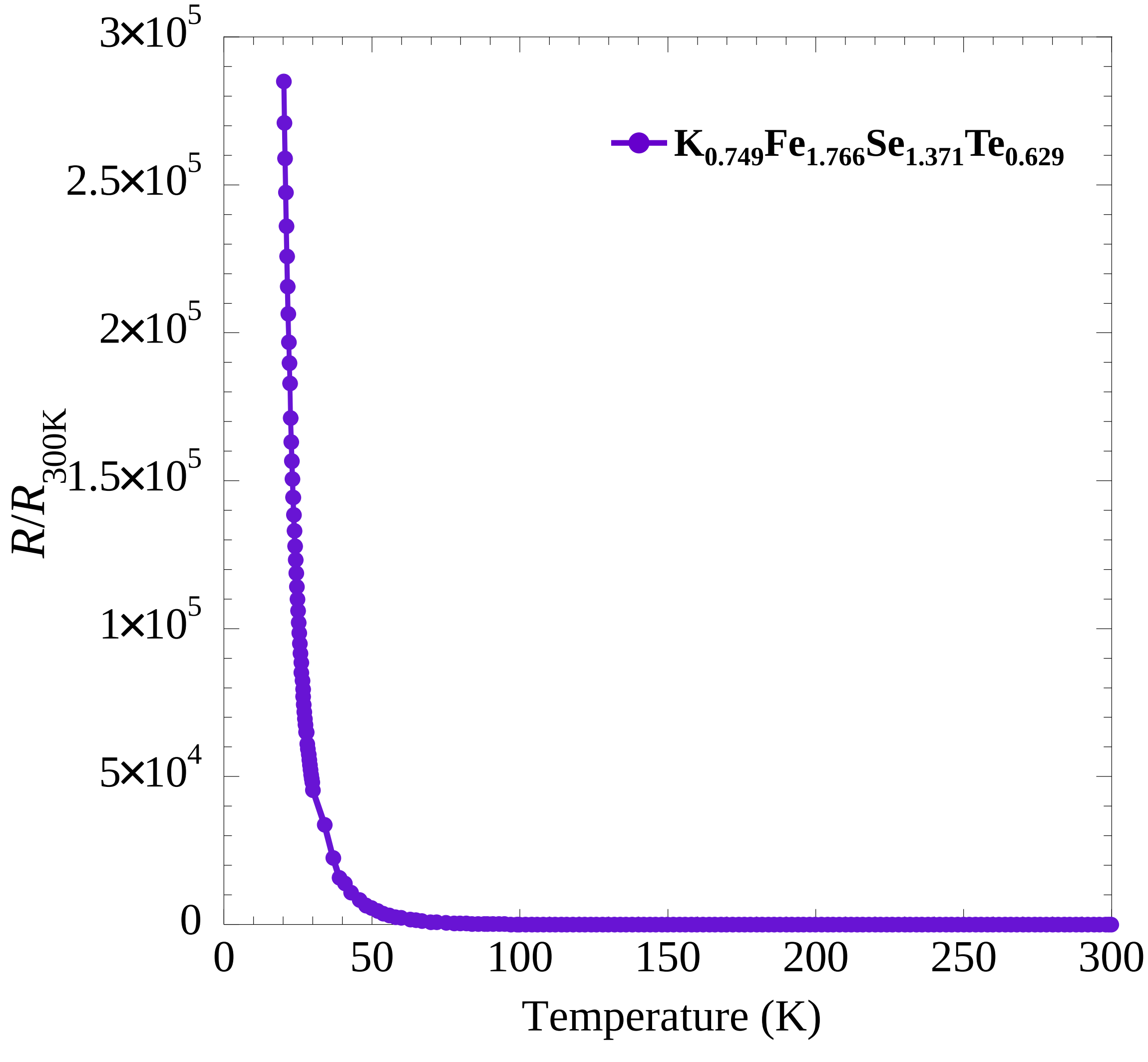

Figure 4

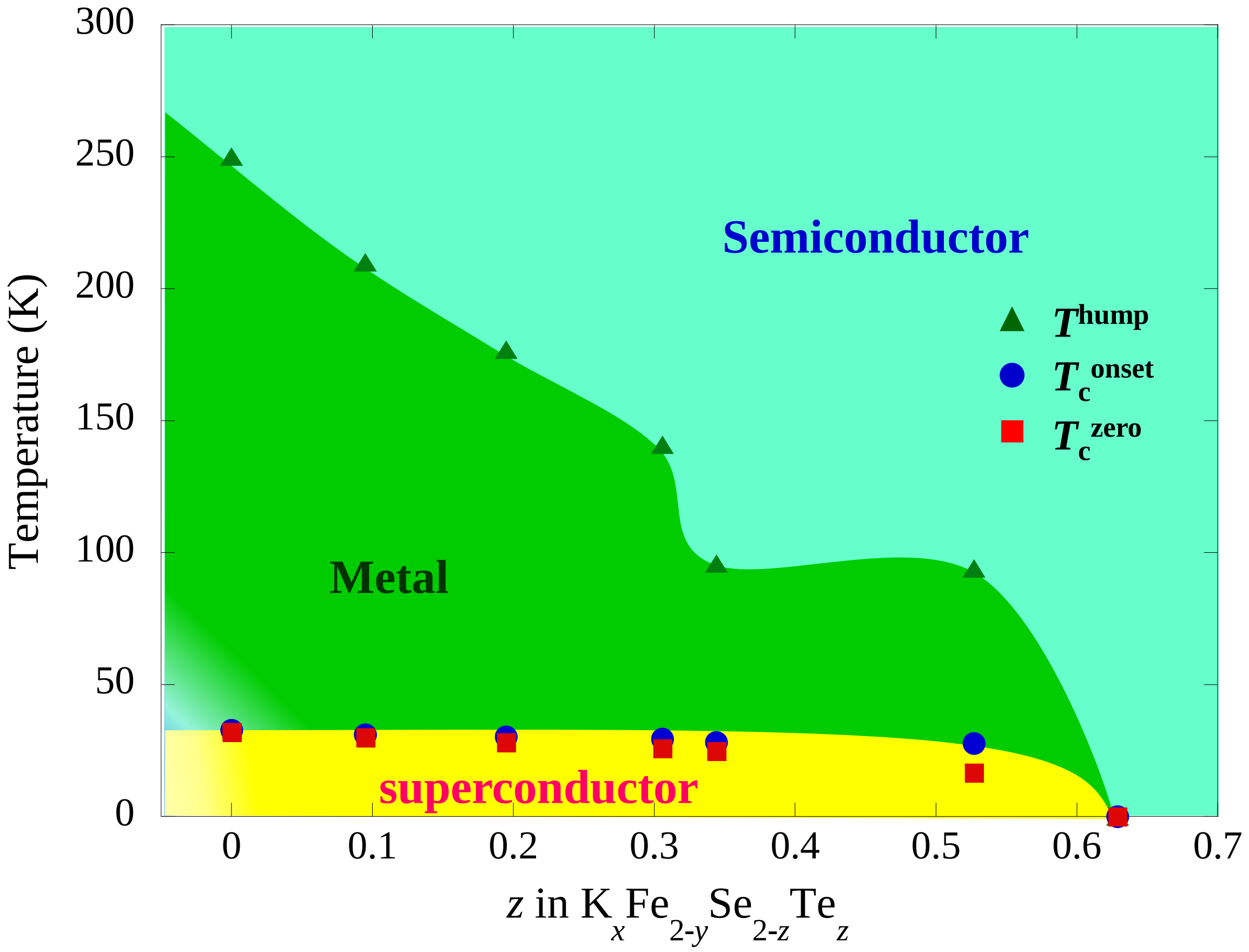

Temperature (K)
300
250
200
150
100
50
0
Semiconductor
Metal
superconductor
T hump
T c onset
T c zero
0
0.1
0.2
0.3
0.4
0.5
0.6
0.7
z in K x Fe 2-y Se 2-z Te z